\documentclass[preprint2]{aastex63}

\usepackage{graphicx}
\usepackage{hhline}
\usepackage{natbib}
\newcommand\edits[1]{#1}

\newcommand{\fermilat}{\emph{Fermi}-LAT} 
\newcommand{\galcoords}[2]{(l, b) = ($#1\degr$, $#2\degr$)} 
\newcommand{\fluxunits}{\ensuremath{ \text{TeV}^{-1}\,\text{cm}^{-2}\,\text{s}^{-1}  }} 
\newcommand{\intfluxunits}{\ensuremath{ \text{cm}^{-2}\,\text{s}^{-1}  }} 
\newcommand{\fluxnorm}[3]{\ensuremath{\left( #1 \pm #2 \right) \times 10^{#3}\, \fluxunits}} 
\newcommand{\statsys}[3]{$#1\pm#2_\mathrm{stat}\pm#3_\mathrm{sys}$} 
\newcommand{\ucla}{Department of Physics and Astronomy, University of California, Los Angeles, CA 90095, USA; \url{mbuchove@ucla.edu}, \url{jryan@astro.ucla.edu}}

\begin{document}

\shorttitle{VERITAS Observations of the Galactic Center Region at Multi-TeV Gamma-Ray Energies}
\title{VERITAS Observations of the Galactic Center Region at Multi-TeV Gamma-Ray Energies}

\shortauthors{Adams et al.}

\author{C.~B.~Adams}\affiliation{Department of Physics and Astronomy, Barnard College, Columbia University, NY 10027, USA}
\author{W.~Benbow}\affiliation{Center for Astrophysics $|$ Harvard \& Smithsonian, Cambridge, MA 02138, USA}
\author{A.~Brill}\affiliation{Physics Department, Columbia University, New York, NY 10027, USA}
\author{R.~Brose}\affiliation{Institute of Physics and Astronomy, University of Potsdam, 14476 Potsdam-Golm, Germany and DESY, Platanenallee 6, 15738 Zeuthen, Germany}
\author{M.~Buchovecky}\affiliation{\ucla}
\author{M.~Capasso}\affiliation{Department of Physics and Astronomy, Barnard College, Columbia University, NY 10027, USA}
\author{J.~L.~Christiansen}\affiliation{Physics Department, California Polytechnic State University, San Luis Obispo, CA 94307, USA}
\author{A.~J.~Chromey}\affiliation{Department of Physics and Astronomy, Iowa State University, Ames, IA 50011, USA}
\author{M.~K.~Daniel}\affiliation{Center for Astrophysics $|$ Harvard \& Smithsonian, Cambridge, MA 02138, USA}
\author{M.~Errando}\affiliation{Department of Physics, Washington University, St. Louis, MO 63130, USA}
\author{A.~Falcone}\affiliation{Department of Astronomy and Astrophysics, 525 Davey Lab, Pennsylvania State University, University Park, PA 16802, USA}
\author{Q.~Feng}\affiliation{Department of Physics and Astronomy, Barnard College, Columbia University, NY 10027, USA}
\author{J.~P.~Finley}\affiliation{Department of Physics and Astronomy, Purdue University, West Lafayette, IN 47907, USA}
\author{L.~Fortson}\affiliation{School of Physics and Astronomy, University of Minnesota, Minneapolis, MN 55455, USA}
\author{A.~Furniss}\affiliation{Department of Physics, California State University - East Bay, Hayward, CA 94542, USA}
\author{A.~Gent}\affiliation{School of Physics and Center for Relativistic Astrophysics, Georgia Institute of Technology, 837 State Street NW, Atlanta, GA 30332-0430}
\author{G.~H.~Gillanders}\affiliation{School of Physics, National University of Ireland Galway, University Road, Galway, Ireland}
\author{C.~Giuri}\affiliation{DESY, Platanenallee 6, 15738 Zeuthen, Germany}
\author{D.~Hanna}\affiliation{Physics Department, McGill University, Montreal, QC H3A 2T8, Canada}
\author{O.~Hervet}\affiliation{Santa Cruz Institute for Particle Physics and Department of Physics, University of California, Santa Cruz, CA 95064, USA}
\author{J.~Holder}\affiliation{Department of Physics and Astronomy and the Bartol Research Institute, University of Delaware, Newark, DE 19716, USA}
\author{G.~Hughes}\affiliation{Center for Astrophysics $|$ Harvard \& Smithsonian, Cambridge, MA 02138, USA}
\author{T.~B.~Humensky}\affiliation{Physics Department, Columbia University, New York, NY 10027, USA}
\author{W.~Jin}\affiliation{Department of Physics and Astronomy, University of Alabama, Tuscaloosa, AL 35487, USA}
\author{P.~Kaaret}\affiliation{Department of Physics and Astronomy, University of Iowa, Van Allen Hall, Iowa City, IA 52242, USA}
\author{N.~Kelley-Hoskins}\affiliation{DESY, Platanenallee 6, 15738 Zeuthen, Germany}
\author{M.~Kertzman}\affiliation{Department of Physics and Astronomy, DePauw University, Greencastle, IN 46135-0037, USA}
\author{D.~Kieda}\affiliation{Department of Physics and Astronomy, University of Utah, Salt Lake City, UT 84112, USA}
\author{F.~Krennrich}\affiliation{Department of Physics and Astronomy, Iowa State University, Ames, IA 50011, USA}
\author{S.~Kumar}\affiliation{Physics Department, McGill University, Montreal, QC H3A 2T8, Canada}
\author{M.~J.~Lang}\affiliation{School of Physics, National University of Ireland Galway, University Road, Galway, Ireland}
\author{M.~Lundy}\affiliation{Physics Department, McGill University, Montreal, QC H3A 2T8, Canada}
\author{G.~Maier}\affiliation{DESY, Platanenallee 6, 15738 Zeuthen, Germany}
\author{P.~Moriarty}\affiliation{School of Physics, National University of Ireland Galway, University Road, Galway, Ireland}
\author{R.~Mukherjee}\affiliation{Department of Physics and Astronomy, Barnard College, Columbia University, NY 10027, USA}
\author{D.~Nieto}\affiliation{Institute of Particle and Cosmos Physics, Universidad Complutense de Madrid, 28040 Madrid, Spain}
\author{M.~Nievas-Rosillo}\affiliation{DESY, Platanenallee 6, 15738 Zeuthen, Germany}
\author{S.~O'Brien}\affiliation{Physics Department, McGill University, Montreal, QC H3A 2T8, Canada}
\author{R.~A.~Ong}\affiliation{\ucla}
\author{A.~N.~Otte}\affiliation{School of Physics and Center for Relativistic Astrophysics, Georgia Institute of Technology, 837 State Street NW, Atlanta, GA 30332-0430}
\author{K.~Pfrang}\affiliation{DESY, Platanenallee 6, 15738 Zeuthen, Germany}
\author{M.~Pohl}\affiliation{Institute of Physics and Astronomy, University of Potsdam, 14476 Potsdam-Golm, Germany and DESY, Platanenallee 6, 15738 Zeuthen, Germany}
\author{R.~R.~Prado}\affiliation{DESY, Platanenallee 6, 15738 Zeuthen, Germany}
\author{E.~Pueschel}\affiliation{DESY, Platanenallee 6, 15738 Zeuthen, Germany}
\author{J.~Quinn}\affiliation{School of Physics, University College Dublin, Belfield, Dublin 4, Ireland}
\author{K.~Ragan}\affiliation{Physics Department, McGill University, Montreal, QC H3A 2T8, Canada}
\author{P.~T.~Reynolds}\affiliation{Department of Physical Sciences, Cork Institute of Technology, Bishopstown, Cork, Ireland}
\author{D.~Ribeiro}\affiliation{Physics Department, Columbia University, New York, NY 10027, USA}
\author{G.~T.~Richards}\affiliation{Department of Physics and Astronomy and the Bartol Research Institute, University of Delaware, Newark, DE 19716, USA}
\author{E.~Roache}\affiliation{Center for Astrophysics $|$ Harvard \& Smithsonian, Cambridge, MA 02138, USA}
\author{J.~L.~Ryan}\affiliation{\ucla}
\author{M.~Santander}\affiliation{Department of Physics and Astronomy, University of Alabama, Tuscaloosa, AL 35487, USA}
\author{S.~Schlenstedt}\affiliation{CTAO, Saupfercheckweg 1, 69117 Heidelberg, Germany}
\author{G.~H.~Sembroski}\affiliation{Department of Physics and Astronomy, Purdue University, West Lafayette, IN 47907, USA}
\author{R.~Shang}\affiliation{\ucla}
\author{B.~Stevenson}\affiliation{\ucla}
\author{S.~P.~Wakely}\affiliation{Enrico Fermi Institute, University of Chicago, Chicago, IL 60637, USA}
\author{A.~Weinstein}\affiliation{Department of Physics and Astronomy, Iowa State University, Ames, IA 50011, USA}
\author{D.~A.~Williams}\affiliation{Santa Cruz Institute for Particle Physics and Department of Physics, University of California, Santa Cruz, CA 95064, USA}

\begin{abstract}
The Galactic Center (GC) region hosts a variety of powerful astronomical sources and rare astrophysical processes that emit a large flux of non-thermal radiation.
The inner 375 pc $\times$ 600 pc region, called the Central Molecular Zone (CMZ), is home to the supermassive black hole Sagittarius A*, massive cloud complexes, and particle accelerators such as supernova remnants. 
We present the results of our improved analysis of the very-high-energy (VHE) gamma-ray emission above 2 TeV from the GC using 125 hours of data taken with the VERITAS imaging-atmospheric Cherenkov telescope between 2010 and 2018.
The central source VER J1745--290, consistent with the position of Sagittarius A*, is detected at a significance of 38 standard deviations above the background level $(38\sigma)$, and we report its spectrum and light curve.
Its differential spectrum is consistent with a power law with exponential cutoff, with a spectral index of $2.12^{+0.22}_{-0.17}$, a flux normalization at 5.3 TeV of $1.27^{+0.22}_{-0.23}\times 10^{-13}$ \fluxunits, and cutoff energy of $10.0^{+4.0}_{-2.0}$ TeV.
We also present results on the diffuse emission near the GC, obtained by combining data from multiple regions along the GC ridge which yield a cumulative significance of $9.5\sigma$.
The diffuse GC ridge spectrum is best fit by a power law with a hard index of 2.19 $\pm$ 0.20, showing no evidence of a cutoff up to 40 TeV.
This strengthens the evidence for a potential accelerator of PeV cosmic rays being present in the GC.
We also provide spectra of the other sources in our field of view with significant detections, composite supernova remnant G0.9+0.1 and HESS J1746--285.
\end{abstract}


\cleardoublepage
\section{Introduction} \label{sec:intro}
The Galactic Center (GC) is host to numerous potential sites of particle acceleration, including the supermassive \citep[$M\sim2.6 \times 10^6 M_\odot$;][]{Schodel2002,Ghez2003,Ghez2005,Gillessen2009} black hole Sagittarius A* (hereafter Sgr A*), supernova remnants (SNRs), and pulsar wind nebulae (PWNe).
The GC also contains dense molecular clouds, constituting the Central Molecular Zone \citep[CMZ][]{Morris1996}.
Very-high-energy (VHE; $>100$\,GeV) gamma-ray emission has been detected from the direction of the GC with imaging-atmospheric Cherenkov telescopes \citep[IACTs;][]{Tsuchiya2004,Kosack2004,Aharonian2004,Albert2006, Archer2014}, leading to important discoveries in high-energy astrophysics and constraints on models for particle dark matter \citep{vanEldik2015}.

Sources of VHE gamma rays include the strong central source VER J1745--290, coincident with both Sgr A* and PWN 359.95--0.04 \citep{Archer2016}, composite SNR G0.9+0.1 \citep{Aharonian2005g}, and an unidentified source \edits{variously identified as VER 1746--289 \citep{Archer2014,Archer2016,Ahnen2017} or HESS J1746--285 \citep{Abdalla2018}.}
There is also diffuse emission that extends along the Galactic plane  \citep{Aharonian2006a,Archer2016,Abramowski2016,Ahnen2017,Abdalla2018}.
The spectrum of VER J1745--290 \edits{(which we consider to be the same object as HESS 1745--290)} has a photon index $\Gamma\sim2.2$, and exhibits a break at $\sim10$\,TeV \citep{Albert2006,Aharonian2009,Archer2016}, while the diffuse emission spectrum has $\Gamma\sim2.3$ with no break or cutoff up to tens of TeV \citep{Aharonian2006c,Abramowski2016}.

The origin of the GC VHE emission remains undetermined, due in part to source confusion and the limitations of current instruments.
The source of VER J1745--290 may be Sgr A* \citep{Atoyan2004,Aharonian2005a,Chernyakova2011,Ballantyne2011,Fatuzzo2012,Kusunose2012,Fujita2017,Rodriguez2019} or PWN G359.95-0.04 \citep{Wang2006,Hinton2007}, with which it is spatially coincident \citep{Acero2010}.
Other possible origins include the annihilation of dark matter particles \citep{Horns2005,Bergstrom2005a,Bergstrom2005b,Profumo2005,Aharonian2006b,Belikov2012,Cembranos2012,Cembranos2013,Gammaldi2016} or a population of millisecond pulsars \citep{Bednarek2013,Bartels2016,Guepin2018}.
The mechanism of gamma-ray emission may be predominantly due to hadronic processes, where relativistic protons interact with gas and subsequently produce gamma rays through neutral pion decay \citep{Aharonian2005a,Chernyakova2011,Ballantyne2011,Fatuzzo2012,Linden2012,Guepin2018}, leptonic processes where gamma rays are produced when electrons and positrons undergo inverse Compton scattering off a radiation field \citep{Atoyan2004,Hinton2007,Kusunose2012,Lacroix2016}, or a combination of processes (hybrid scenario), where leptons produce high energy, but not VHE, gamma rays \citep{Guo2013}.
Both the correlation of VHE emission with the CMZ and the lack of a cutoff in the diffuse spectrum support a hadronic scenario, capable of explaining both VER J1745--290 and the diffuse emission \citep{Aharonian2006c,Linden2012,Abramowski2016}.
Measurement of the diffuse spectrum by H.E.S.S. up to energies of tens of TeV with no evidence of a cutoff has also been interpreted as evidence for the existence of PeV protons within the central 10 parsecs of the Galactic Center, accelerated by Sgr A* \citep{Abramowski2016}.
While cosmic rays are known to extend up to PeV energies \citep[e.g.][]{Horandel2003}, few, if any, accelerators of PeV cosmic rays, or `PeVatrons', have been clearly established \citep[e.g.][]{Abramowski2016,Abeysekara2020}.
Discovering the nature of PeVatrons in our Galaxy is thus a particularly important step in understanding the origins of cosmic rays.

In addition to the spectra and morphology of the astrophysical sources detected, their variability also constrains models of VHE emission.
Correlated variability in different wavebands would suggest a common origin, while variability timescales can constrain the nature of the acceleration mechanism \citep{Ballantyne2011} or the size of the emission region.
To date, no variability has been detected in the TeV emission from the direction of Sgr A* \citep{Albert2006,Aharonian2009,Ahnen2017}, suggesting a differing origin of the VHE radiation from the variable IR and X-ray emission \citep{Wang2006}.

In the neighboring high-energy waveband, an excess at the GC peaking near 3 GeV has been detected by \fermilat{} \citep{Atwood2009,Ackermann2017}.
Proposed explanations include
unresolved point sources \citep[e.g.][]{Abazajian2011,Brandt2015,Bartels2016,Lee2016,Macias2018,Buschmann2020},
annihilation of dark matter particles \citep[e.g.][]{Hooper2011,Daylan2016,Leane2019},
or injection of electrons or protons into the interstellar medium \citep[e.g.][]{Chernyakova2011,Petrovic2014,Carlson2014,Gaggero2015,Malyshev2015}.
Certain models have signatures at TeV energies as well \citep[e.g.][]{Yusef-Zadeh2013}.

In this work, we expand upon previous VERITAS analyses of the Galactic Center using additional data, taken between 2010 and 2018, and using improved analysis techniques.
We extend the measured spectrum of the central source VER J1745--290 to 40 TeV, and present new limits on its VHE variability.
We report the first VERITAS spectrum of the diffuse Galactic ridge emission spanning $-0.7\degr$ to $+1.3\degr$ in galactic longitude.
We also provide spectra and positions for G0.9+0.1 and HESS J1746--285.

\section{Data and Methodology}
\label{sec:methods}

\subsection{Overview of GC Observations and Analysis}
\label{sec:overview}
The Very Energetic Radiation Imaging Telescope Array System (VERITAS) is a ground-based array of four 12-meter IACTs located at the Fred Lawrence Whipple Observatory in Amado, Arizona at coordinates $31\arcdeg\,40\arcmin,30\arcsec$\,N, $110\arcdeg\,57\arcmin\,07\arcsec$\,W and at an altitude of 1268 m above sea level.
Between 2010 April and 2018 June, VERITAS accumulated 155 hours of data-quality assessed observations of the GC region. After time cuts and dead-time corrections are applied, the total exposure is approximately 125 hours; an additional 40 hours compared to \citet{Archer2016}.

Due to the location of VERITAS, the GC can only be observed at large zenith angles (LZA) between 59 and 66 degrees.
Observing at LZA increases an IACT's effective area and minimum energy threshold, compared to smaller zenith angles \citep{Sommers1987,Konopelko1999}.
The thicker atmosphere along the line of sight results in the gamma-ray-initiated electromagnetic particle cascades (``air showers'') originating further away from the telescopes, on average.
The air shower images thus appear fainter due to the light spreading out as well as additional atmospheric attenuation, raising the energy threshold to about 2 TeV for the GC analysis.
The larger distance and zenith angle of LZA air showers also result in larger Cherenkov light pools, increasing the effective area at $60\degr$ to about four times greater than the effective area at a smaller zenith angle of $20\degr$ for energies above 10 TeV.
LZA observations also introduce various biases into gamma ray reconstruction and require specialized analysis methods to correct them.

To estimate the arrival directions of air showers, we use an improved version of the \textit{displacement} method \citep[e.g.][]{Akerlof1991,Buckley1998,Lessard2001,Senturk2011}.
The standard \emph{geometric} method \citep[e.g.][]{Hofmann1999,Krawczynski2006}, which uses the intersection of the air shower images' major axes recorded in each telescope, performs poorly at zenith angles greater than $40\degr$ \citep{Archer2014}.
At LZA, the telescope separation in the plane perpendicular to the shower axis shrinks due to projection effects, and the larger effective area results in greater impact distances between the shower core and telescopes on average.
Both of these effects reduce the viewing angle differences of the telescopes, causing the major axes to be closer to parallel, thereby increasing the uncertainty in the estimated arrival direction.
The \textit{displacement} method does not rely on stereo information, and instead estimates the distance between the source position and a shower image centroid (the \textit{disp} parameter) based on Monte-Carlo simulations of gamma-ray showers.
Various implementations of the \textit{displacement} method have been applied to LZA observations of the GC previously \citep{Kosack2004,Archer2014,Archer2016}.
Our new method uses the same simulations and image parametrization as the standard VERITAS \textit{displacement} method implementation \citep{Aliu2012,Archer2014}, but uses boosted decision trees (BDTs) to estimate the \textit{disp} parameter instead of lookup tables.
We use BDTs with gradient boosting, as implemented in the Toolkit for MultiVariate data Analysis \citep{Hoecker2007} with ROOT \citep{Antcheva2009}.

BDTs have also been shown to improve energy reconstruction \citep{Albert2008} and gamma-hadron separation power \citep{Ohm2009, Krause2017, HGPS} for IACT data.
We use the BDT method for energy reconstruction as well, and find that it reduces the energy bias and improves the energy resolution by approximately 10\% in LZA analyses.
The results were validated on Crab Nebula data to ensure the spectrum was consistent with expectations \citep{Buchovecky2019}.

\subsection{Gamma-Hadron Separation}
IACTs observe optical Cherenkov radiation from air showers initiated by both gamma rays and hadrons.
Instrument sensitivity benefits from the ability to distinguish gamma-ray showers from the substantial background of hadronic showers, in a process termed ``gamma-hadron separation'' \citep[e.g.][]{Krawczynski2006}.
The standard method of gamma-hadron separation compares the values of an event's mean-scaled stereo Hillas parameters \citep{Hillas1985} to fixed parameter ranges (``cuts''), outside of which events are considered cosmic-ray-like and are excluded from the analysis.
LZA observations result in distributions of stereo parameters that differ in shape from standard observations for which the standard optimal fixed cut values were found. 
To improve gamma-hadron separation, optimal cut values that maximize sensitivity for LZA data are found, using 20 hours of data taken on the Crab Nebula with a distribution of zenith angles that closely matches the GC dataset.
The statistical significance of the Crab Nebula was found using the ring-background method \citep[RBM;][]{Berge2006}, varying the cut values for mean-scaled width (MSW), mean-scaled length (MSL), and size.
The optimized LZA fixed cuts were then validated on a separate 25 hours of LZA Crab Nebula data \citep{Buchovecky2019}.
We find that the LZA-optimized cuts provide a roughly 10\% gain in sensitivity, compared to the standard cuts.

\subsection{Acceptance}
The acceptance, or relative detection efficiency, is approximately radially symmetric across the VERITAS field of view.
Since the acceptance is non-uniform, acceptance-weighted exposure times are used in calculations of gamma-ray excess and significance \citep{Li1983,Spengler2015}.
The radial acceptance function depends only on the angular distance between the tracking position of the telescope pointing and the reconstructed event position.
A zenith acceptance correction is also applied, following the procedure described in \citep{Zitzer2017}.
This zenith correction to the acceptance map is especially important at large zenith angles, where deviations from the uncorrected radially symmetric map can exceed 10\%.

\subsection{Source Localization}
Locations of point sources are determined by fitting a two-dimensional model of the VERITAS point spread function (PSF) plus a constant to an acceptance-corrected excess map.
The PSF at LZA is better described by the radial King function \citep{Read2011} than a two-dimensional Gaussian.
The King function is given by
\begin{equation} \label{eqn:king_function}
K\left(r,\sigma,\gamma\right)=\frac{1}{2\pi\sigma^{2}}\left(1-\frac{1}{\gamma}\right)\left[1+\frac{1}{2\gamma}\cdot\frac{r^2}{\sigma^{2}}\right]^{-\gamma},
\end{equation}
where $K$ is the probability density of the reconstructed event position, $r$ is the radial distance from the source position, and $\sigma$ and $\gamma$ are free parameters.
Values of $\sigma$ and $\gamma$ are derived from simulated point-source data, yielding $\sigma=0.056 \degr$ and $\gamma=2.3$.
The resulting 68\% containment radius of the PSF model is $0.124\degr$.
In our analysis of the data, position fits are conducted in the Galactic coordinate plane, and are performed out to a radius of $0.5\degr$.

\subsection{Spectral Models}
Differential flux spectra are calculated in discrete energy bins for a $0.1\degr$ region around the source using the reflected region method \citep[RRM;][]{Aharonian2001} for background estimation.
Power law (PL), exponentially cutoff power law (ECPL), and smoothly broken power law (BPL) fits to the binned spectral points are explored.
The PL model is defined as
\begin{equation} \label{eq:pure_pow_law}
\frac{dN}{dE} = N_0 \left(\frac{E}{E_0}\right)^{-\Gamma},
\end{equation} 
where $\Gamma$ is the differential spectral index and $N_0$ is the flux normalization at energy $E_0$.
The ECPL is defined as
\begin{equation} \label{eq:ecpl}
\frac{dN}{dE} = N_0 \left(\frac{E}{E_0}\right)^{-\Gamma}\times \exp{\left(-\frac{E}{E_\mathrm{cut}}\right)},
\end{equation}
where the additional parameter $E_\mathrm{cut}$ is the cutoff energy.
The BPL is defined as
\begin{equation} \label{eq:bpl}
\frac{dN}{dE} = N_0 \frac{(E/E_0)^{-\Gamma_1}}{1+\left(E/E_\mathrm{break}\right)^{\Gamma_2-\Gamma_1}},
\end{equation}
where $\Gamma_1$ and $\Gamma_2$ are the spectral indices before and after the break energy $E_\mathrm{break}$.

\section{Results} \label{sec:results}

\subsection{Significance Map of the Galactic Center}  \label{sec:results_skymap}
\begin{figure*}[ht]
\centering
\includegraphics[width=0.95\textwidth]{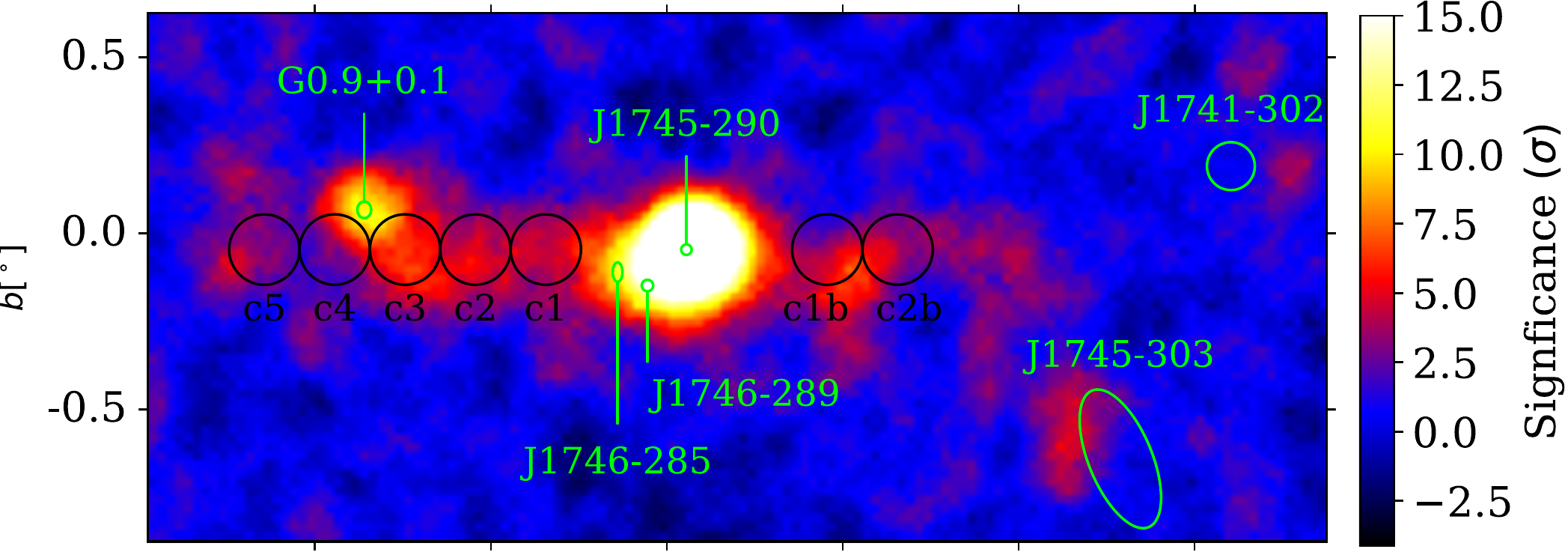}
\includegraphics[width=0.95\textwidth]{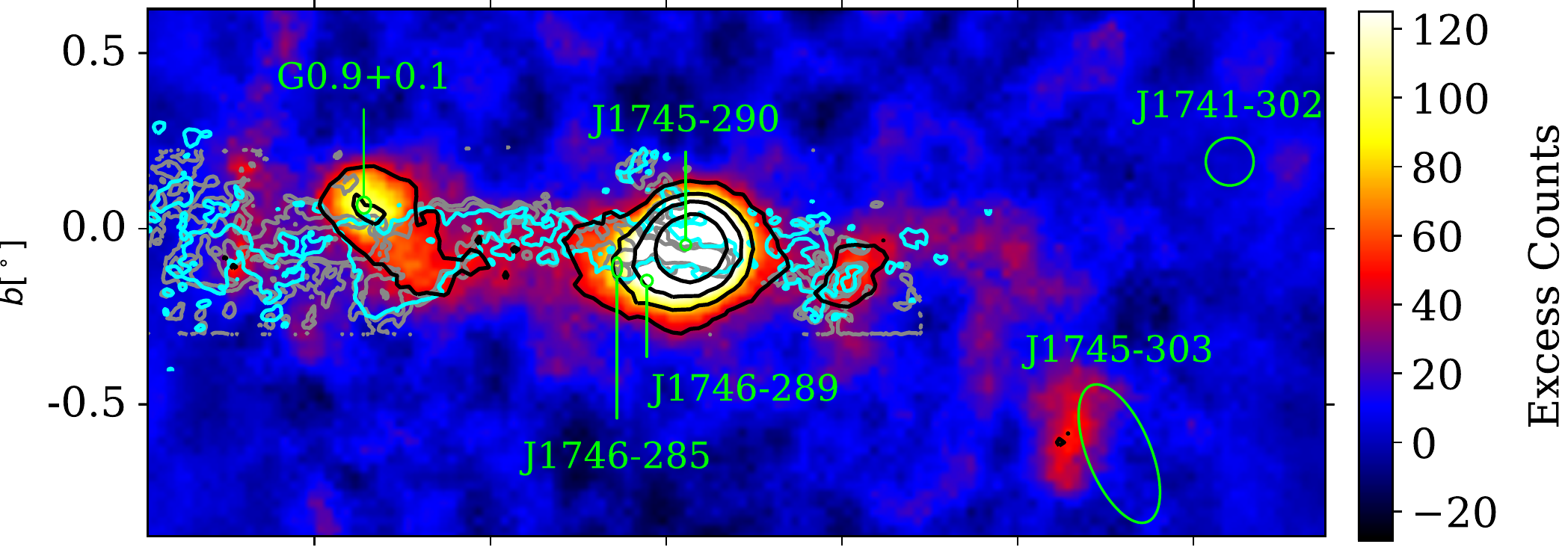}
\includegraphics[width=0.95\textwidth]{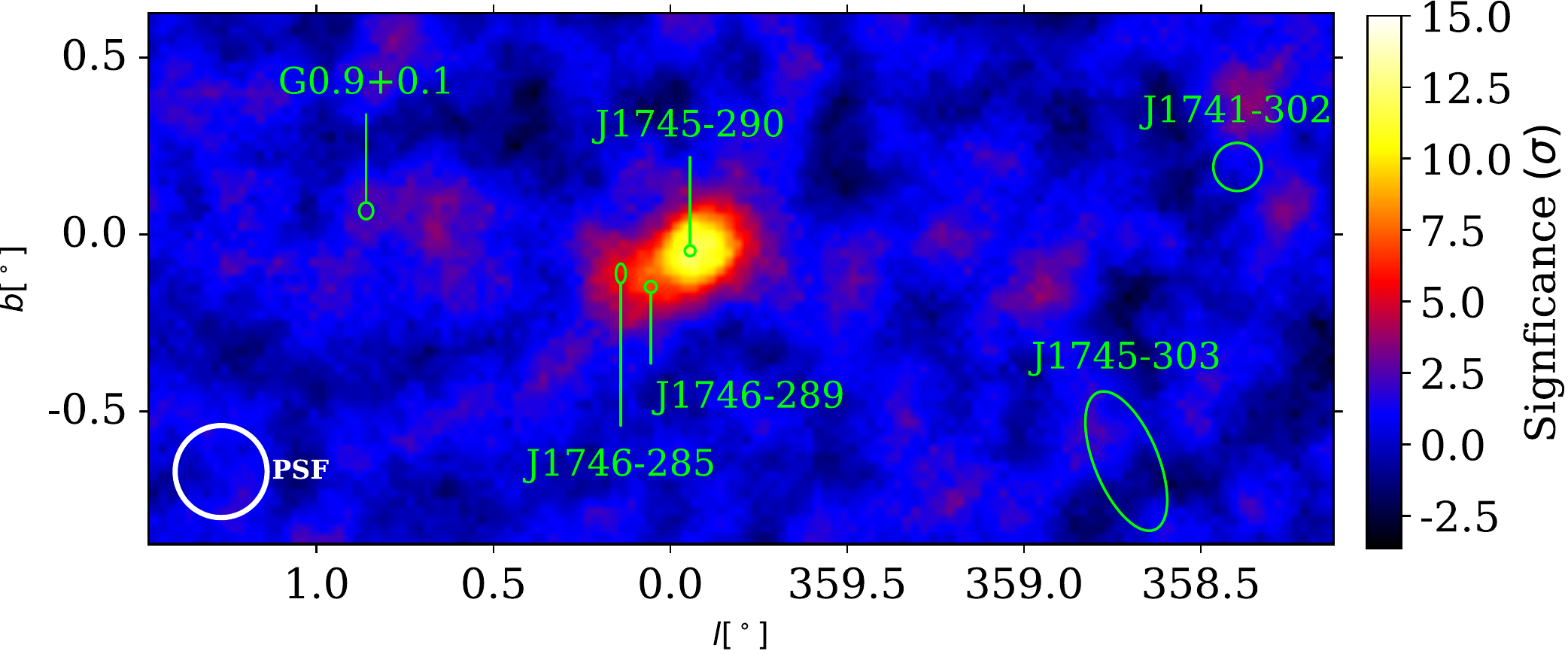}
  \caption{Maps of the statistical significance for gamma-ray-like events detected by VERITAS above 2 TeV (top) and 10 TeV (bottom) from this analysis, as well as the map of acceptance-corrected correlated excess counts above 2 TeV (middle).
  The significance scale is limited to $15\sigma$ so the structure along the ridge can be seen in detail.
  \edits{Each pixel of} the correlated excess map displays the excess counts integrated over \edits{$0.1\degr$ circular signal regions centered on the pixel,} and has been convolved with the VERITAS PSF, shown in the bottom-left.
  Positions and 68\% confidence regions are shown for previously detected point sources (green ellipses),
  while the ellipses for J1741--302 and J1745--303 represent their spatial extents \citep{Abdalla2018b,Aharonian2006a}.
  The signal regions used in the diffuse ridge analysis are shown in the top panel (black circles), labeled as in \citet{Abramowski2016}.
  Contours of CS \citep[cyan;][]{Tsuboi1999} and HCCCN line emission \citep[gray;][]{Jones2011}, which trace dense molecular gas, are shown in the middle panel, along with significance contours of 5, 10, 15, and $25\sigma$ (black) from this analysis.}
  \label{fig:gc_sigmap_full_fov}
\end{figure*}

The map of statistical significance for events above 2 TeV of the inner $3\degr\times1.25\degr$ of the GC region is shown in the top panel of \autoref{fig:gc_sigmap_full_fov}.
The positional uncertainties of HESS J1745--290, G0.9+0.1, VER J1746--289, and HESS J1746--285 are indicated.
The color scale, which represents statistical significance, is limited to $15\sigma$ because the peak significance is over $39\sigma$ and would obscure the remaining structure of diffuse emission.
Strong diffuse emission is visible along the Galactic ridge, extending about $1.25\degr$ in each direction from the center.
The map of excess gamma-ray counts (corrected for acceptance) detected by VERITAS is shown in the middle panel of \autoref{fig:gc_sigmap_full_fov}.
Contours of CS $J=1-0$ line emission and HCCCN emission, which trace dense gas, are also shown.
There is an appearance of a moderate correlation between the diffuse emission and dense gas.

The significance map for events above 10 TeV is shown in the bottom of \autoref{fig:gc_sigmap_full_fov}.
Aside from the central emission, the strongest signal comes from HESS J1746--285.

\begin{figure}[!t]
 \centering
  \includegraphics[width=0.48\textwidth]{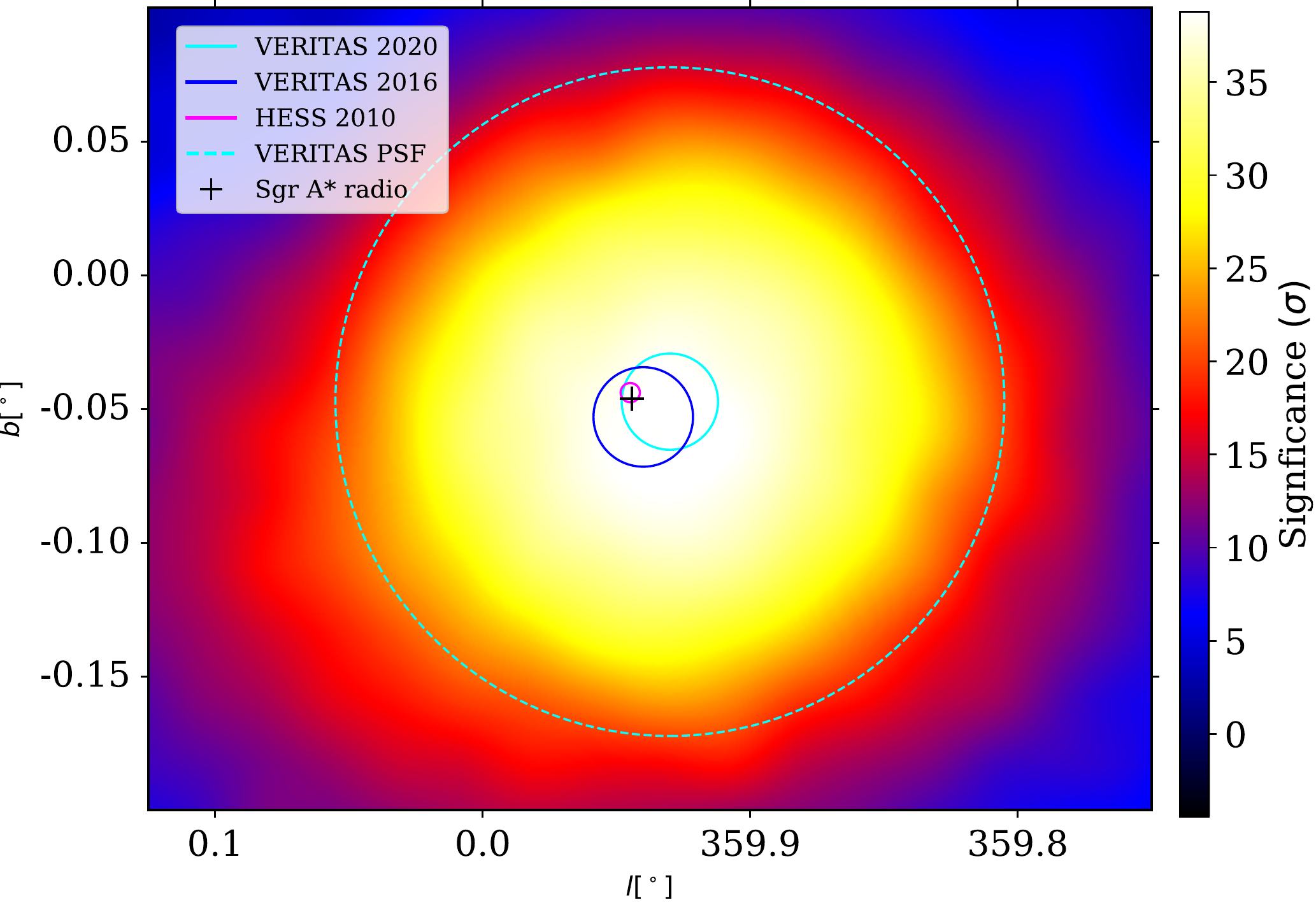}
  \caption{Significance map of the central source VER J1745--290, with the full scale of statistical significance.
The 68\% confidence region on the position of the source is given for this analysis (cyan) and \citet[][blue]{Archer2016}, \edits{as well as that of \citet{Acero2010} for HESS J1745--290 (magenta).}
The dashed line is the 68\% containment radius of the VERITAS PSF, centered on the best-fit position of VER J1745--290 from this analysis.
The radio position of Sgr A* from \citet{Reid2004} is marked by the black cross.}
  \label{fig:gc_sigmap_zoomed}
\end{figure}

\begin{figure*}[!t]
 \centering
  \includegraphics[width=0.95\textwidth]{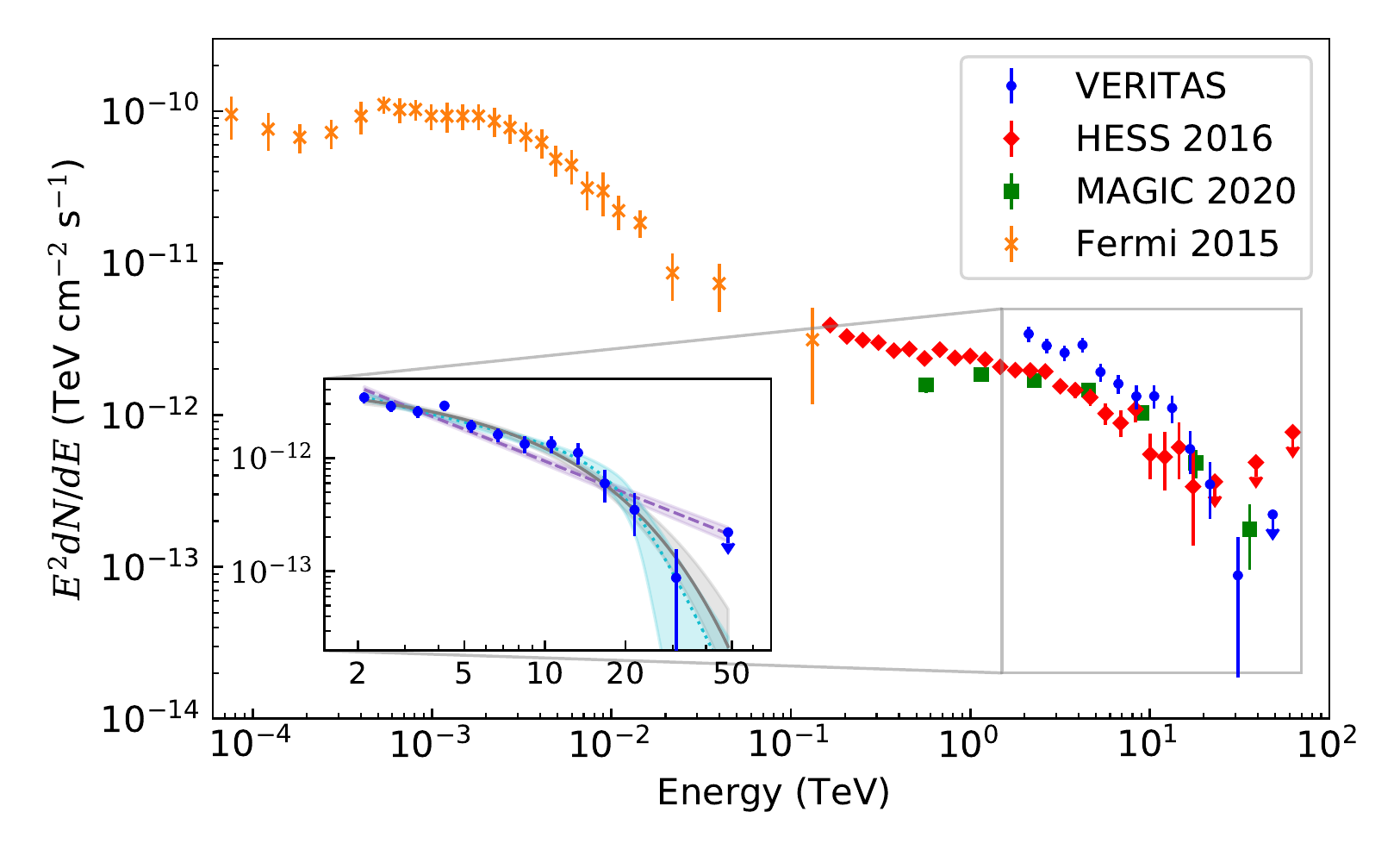}
  \caption
  {\edits{Differential energy spectrum of the central source coincident with the position of Sgr A*, as measured by VERITAS (blue), H.E.S.S. \citep[red;][]{Abramowski2016}, MAGIC \citep[green;][]{Acciari2020}, and \fermilat{} \citep[orange;][]{Malyshev2015}.}
  Error bars represent $1\sigma$ uncertainties in the flux.
  The downward arrows represent 95\% upper limits.
  \edits{In the inset, the best-fit exponentially cutoff power law (gray solid line), broken power law (cyan dotted line), and power law (purple dashed line) are shown.}
  Shaded regions represent the $1\sigma$ confidence band on the model fits.
\label{fig:j1745-290_spectra_ecpl} }
\end{figure*}

\section{VER J1745--290}
A significance map showing the central emission around VER J1745--290 in greater detail is shown in \autoref{fig:gc_sigmap_zoomed}.
The RBM analysis of VER\,J1745--290 yields a statistical significance of $37.5\sigma$, compared to the previous result of $25\sigma$ found in \citet{Archer2016}.
A total excess of 426 gamma-ray-like events above 2 TeV was detected in the $0.1\degr$ circular signal region centered on the nominal position of Sgr A*, \galcoords{359.944}{-0.0462} \citep{Reid2004}.

The best-fit position of VER J1745--290 is \galcoords{359.930}{-0.047}, with uncertainties in $l$ and $b$ of $0.018\degr$.
The uncertainty is dominated by a systematic uncertainty of $0.013\degr$, which is larger at LZA than at higher elevations \citep{Archer2016}.
This position is consistent with the previous position of VER J1745--290 \citep{Archer2016}, the H.E.S.S. position of HESS J1745--290 from \citet{Acero2010}, and the radio position of Sgr A*.
The approximate distance of G359.95--0.04's tail from Sgr A* is about $4\arcsec$, and the mean position of G359.95--0.04 is slightly farther away, as its head is about $8.7\arcsec$ away from Sgr A* \citep{Wang2006}.
Excluding either source as a candidate for the central emission requires a smaller angular uncertainty, which may be achieved by CTA in the future with a point-source localization accuracy of less than $3\arcsec$ \citep{Acharya2018}.

\subsection{Spectrum}
The differential energy spectrum of VER J1745--290 is shown in Figure \ref{fig:j1745-290_spectra_ecpl}.
\edits{Also shown for comparison are the differential energy spectra of the central source, coincident with the position of Sgr A*, as measured by H.E.S.S. \citep{Abramowski2016}, MAGIC \citep{Acciari2020}, and \fermilat{} \citep{Malyshev2015}.}
The binned spectral data are given in Table \ref{tab:spec_points_j1745}.
The events were extracted from the same signal region used in the RBM analysis and include a small amount of contamination from the diffuse emission and HESS J1746--285.
The energy threshold for this analysis is 1.9 TeV, 
 and the highest significant energy bin has an upper edge of 39.8 TeV.
Fit parameters and statistical uncertainties for all functions are included in Table \ref{tab:spectral_fits_sgra}, as well as the chi-squared value and degrees of freedom ($\chi^2$/d.o.f) for each fit.
The systematic uncertainties for the flux normalization and power law index are conservatively estimated to be about 40\% \citep{Archer2016}.
The ECPL and BPL provide adequate fits to the data based on their $p$-values, while the PL does not.
Since the BPL does not provide a substantially better fit than the ECPL with an additional parameter, we consider the ECPL to provide the best spectral fit.
The ECPL fit gives a harder spectral index of $2.12^{+0.22}_{-0.17}$ below the cutoff, compared to the BPL index of $2.59^{+0.14}_{-0.09}$.
The flux normalization of the ECPL, $12.7^{+2.2}_{-2.3}\times 10^{-14}$ \fluxunits, is also higher than that of the BPL, $7.15^{+0.41}_{-0.54}\times 10^{-14}$ \fluxunits.
The cutoff energy of the ECPL, $10.0^{+4.0}_{-2.0}$ TeV, is consistent with the break energy of the BPL, $6.9^{+9.3}_{-1.1}$ TeV.
\edits{The ECPL spectral index and cutoff energy are consistent with H.E.S.S. \citep{Abramowski2016} and MAGIC \citep{Acciari2020} measurements,} while our flux normalization is slightly higher, though not significantly when systematic uncertainties are accounted for.
The decorrelation energy, the energy at which the correlation between the flux normalization and spectral index parameters is minimized, is 5.3 TeV.

The \fermilat{} spectrum is best-fit by a BPL with a break energy at around 2 GeV. 
The index after the break energy, $2.68 \pm 0.05$, is steeper than the spectrum at very high energies,
suggesting that the emissions at GeV and TeV energies are produced by different mechanisms or different populations of particles.

The integral flux above 1 TeV assuming the ECPL model is $2.9\times10^{-12}$ \intfluxunits, which is roughly 10\% of the Crab Nebula flux in the same energy range.

Using a distance from the GC to the Earth of 7.86 kpc \citep{Boehle2016}, the total VHE luminosity above 1 TeV is approximately $8.1\times 10^{34}$ erg s$^{-1}$.
This is consistent with the lower end of the estimate in \citet{Genzel2010}.

\begin{deluxetable*}{|c|c|c|c|c|c|c|}
\tablewidth{0pt}
\tablecaption{
The VER J1745--290 spectrum's best-fit parameters for the power law (PL, \autoref{eq:pure_pow_law}), exponentially cutoff PL (ECPL, \autoref{eq:ecpl}), and smoothly broken PL (BPL, \autoref{eq:bpl}).
Parameters are defined in the text.
The normalization energy for all three fits is $E_0 = 5.3$ TeV.
68\% confidence intervals on the parameters are indicated.
The ECPL and BPL provide adequate fits to the data, based on the $p$-value.
}
\label{tab:spectral_fits_sgra}
\tablehead{ \colhead{Models} & \colhead{$N_0$ ($10^{-14}$ \fluxunits)} & \colhead{$\Gamma_1$} & \colhead{$\Gamma_2$} & \colhead{$E_\mathrm{cut,break}$ (TeV)} & \colhead{$\chi^2$/d.o.f.} & \colhead{$p$-value}
}
\startdata
PL	&	$6.05^{+0.27}_{-0.30}$	&	$2.94^{+0.06}_{-0.05}$	&	N/A 	&	N/A	&	31.9/10	&	0.00042	\\ \hline
ECPL	&	$12.7^{+2.2}_{-2.3}$	&	$2.12^{+0.22}_{-0.17}$	&	N/A 	&	$10.0^{+4.0}_{-2.0}$	&	6.84/9	&	0.65	\\ \hline
BPL	&	$7.15^{+0.41}_{-0.54}$	&	$2.59^{+0.14}_{-0.09}$	&	$19.5^{+3.7}_{-2.7}$	&	$6.9^{+9.3}_{-1.1}$	&	5.68/8	&	0.68	\\
\enddata
\end{deluxetable*}

\subsection{Flux Variability} 
\label{sec:variability_results}
\begin{figure}[!t]
\centering
  \includegraphics[width=0.45\textwidth]{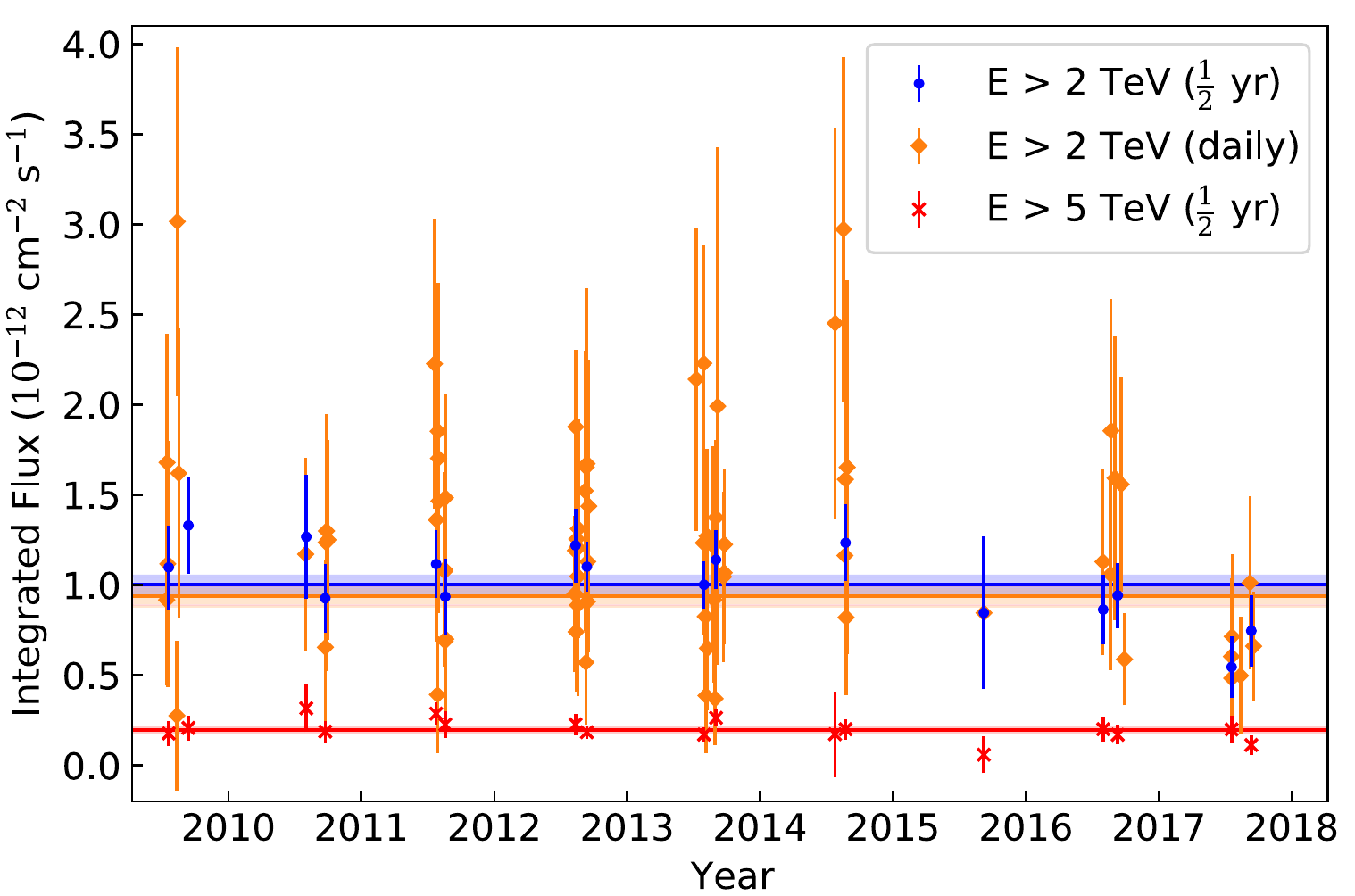}
  \caption{\edits{Semi-annually binned light curves of the integral flux above 2 TeV (blue) and 5 TeV (red) of J1745--290, showing flux versus time.
  Also shown is the daily light curve above 2 TeV (orange).}
  The weighted means are shown as horizontal lines of the corresponding color, while the shaded regions indicate the 68\% confidence interval on the mean.}
  \label{fig:j1745-290_lightcurve}
\end{figure}

Light curves of the integral flux above 2 TeV and 5 TeV of J1745--290 are shown in Figure \ref{fig:j1745-290_lightcurve}.
\edits{The semi-annually binned light curves are fluxes in time bins effectively 1.5 months in duration, corresponding to approximately half of the VERITAS observing season of the GC.
A daily binned light curve of the integral flux above 2 TeV is also shown.
Bins with low statistics, defined as fewer than 4 counts in either the signal or background region, are excluded.
}

\edits{The best-fit means of the light curves are calculated as the uncertainty-weighted means of the flux points.
The best-fit mean of the semi-annually binned light curve of flux above $2$\,TeV is $(1.01 \pm 0.05)\times 10^{-12}$ \intfluxunits.
The fit's $\chi^2$/d.o.f. is 16.64/15, corresponding to a $p$-value of 0.34, consistent with a constant flux hypothesis.
For the semi-annually binned light curve of flux above $5$\,TeV, the best-fit mean is $(1.86\pm0.14)\times 10^{-13}$ \intfluxunits\, with a $\chi^2$/d.o.f. is 20.94/15, corresponding to a $p$-value of 0.14, also consistent with a constant flux.
For the daily light curve of flux above $2$\,TeV, the best-fit mean is $(0.94\pm0.06)\times 10^{-12}$ \intfluxunits, with a $\chi^2$/d.o.f. of 70.1/71, corresponding to a $p$-value of 0.51, consistent with a constant flux.
A typical night of data has a one-hour exposure, and produces a 95\% flux upper limit of $\sim1.5 \times 10^{-12}$ \intfluxunits, above 2\,TeV.
}
The light curve shows no sign of long-term flaring or variability on month timescales, for either energy range.
The apparent decrease in fluxes in the last several time bins, while not statistically significant, may be due to the degradation of the VERITAS mirrors and photodetectors, which will be addressed in a forthcoming publication.

Emission models in which gamma rays are emitted close to the event horizon of Sgr A* predict variability on timescales ranging from minutes to hours \citep[e.g.][]{Aharonian2005a}.
To investigate variability on the timescale of a day, a light curve was also calculated with daily bins.
The $\chi^2$/d.o.f. for this fit is 42.3/52, corresponding to a $p$-value of 0.83, consistent with no variability.

The lack of flux variability is consistent with leptonic models of emission \citep[e.g.][]{Atoyan2004}, as well as extended hadronic models \citep{Aharonian2005b}. This result is also compatible with models that predict variability on timescales longer than a few years.

\section{Analysis of Diffuse Ridge Emission} \label{sec:results_diffuse}
The analysis of the diffuse emission from the GC ridge uses multiple signal regions along the Galactic plane.
The regions used are the same as the seven circular regions used by \citet{Abramowski2016}, and are shown in \autoref{fig:gc_sigmap_full_fov}.
Regions are first analyzed individually using the RRM to get the number of events in the signal and background regions, as well as event energies, exposure times, effective areas, and acceptances.
Exclusion regions are devised for each telescope pointing direction to prevent background regions from being used by more than one signal region.
The results from all regions are then combined to calculate the significance and spectrum.
Observations where the telescope pointing direction is more than $1.5\degr$ from the signal region are excluded from the analysis of that region, due to the large relative uncertainty of the effective area at such offsets.

The map of excess counts in the CMZ from this analysis can be seen in the middle panel of \autoref{fig:gc_sigmap_full_fov}.
The cumulative statistical significance of excess signal after combining the data from \edits{the seven circular regions} is $9.5\sigma$.
An overlay of radio contours from CS line emission \citep{Tsuboi1999} and HCCCN emission \citep{Jones2011} are also shown in \autoref{fig:gc_sigmap_full_fov};
these are two of the most effective known tracers of dense gas material and have a visible correlation with the diffuse gamma-ray emission in the region.

\subsection{Spectrum} \label{sec:results_diffuse_spectrum}
\edits{
The binned spectral data for the differential energy spectrum of the diffuse ridge emission are given in Table \ref{tab:spec_points_diffuse}.
The best fit to our data is a PL with spectral index $2.19 \pm 0.20$ and flux normalization \fluxnorm{3.44}{0.62}{-14} at 5.3 TeV.
The PL provides a good fit, with a $\chi^2$/d.o.f. of 6.03/5, corresponding to a $p$-value of 0.30.
While no significant spectral cut-off is found, consistent with \citet{Abdalla2018}, the 95\% confidence level lower bound on the ECPL's cutoff energy, assuming a spectral index fixed at $2.22$, is 10.3 TeV, consistent with the cutoff observed by \citet{Acciari2020}.
Letting the spectral index vary in the cutoff energy upper limit calculation results in unphysically hard spectral indices.
This calculation neglects the effect of pair absorption, which would presumably shift the cutoff energy higher \citep{Porter2018}.
}

\edits{The point sources contaminate some of the diffuse signal regions.
To estimate this contribution, we model the 2D excess count map, using King functions for the point sources, and using for the diffuse ridge emission the velocity-integrated CS map times a 2D Gaussian of $1.1\degr$ width centered at the GC, as in \citet{Abdalla2018}, convolved with the VERITAS PSF.
The amplitudes of each source component are derived simultaneously from a fit to the correlated excess count map.
Model counts can then be summed in the relevant signal regions for a source, with the fraction of counts coming from other sources constituting the estimated contamination level.
We find that the point sources VER J1745--290, G0.9+0.1, and HESS J1746--285 have an estimated contribution to our total integrated flux measurement of the diffuse ridge emission of approximately 10\%.}

\begin{figure*}[!t]
 \centering
  \includegraphics[width=0.95\textwidth]{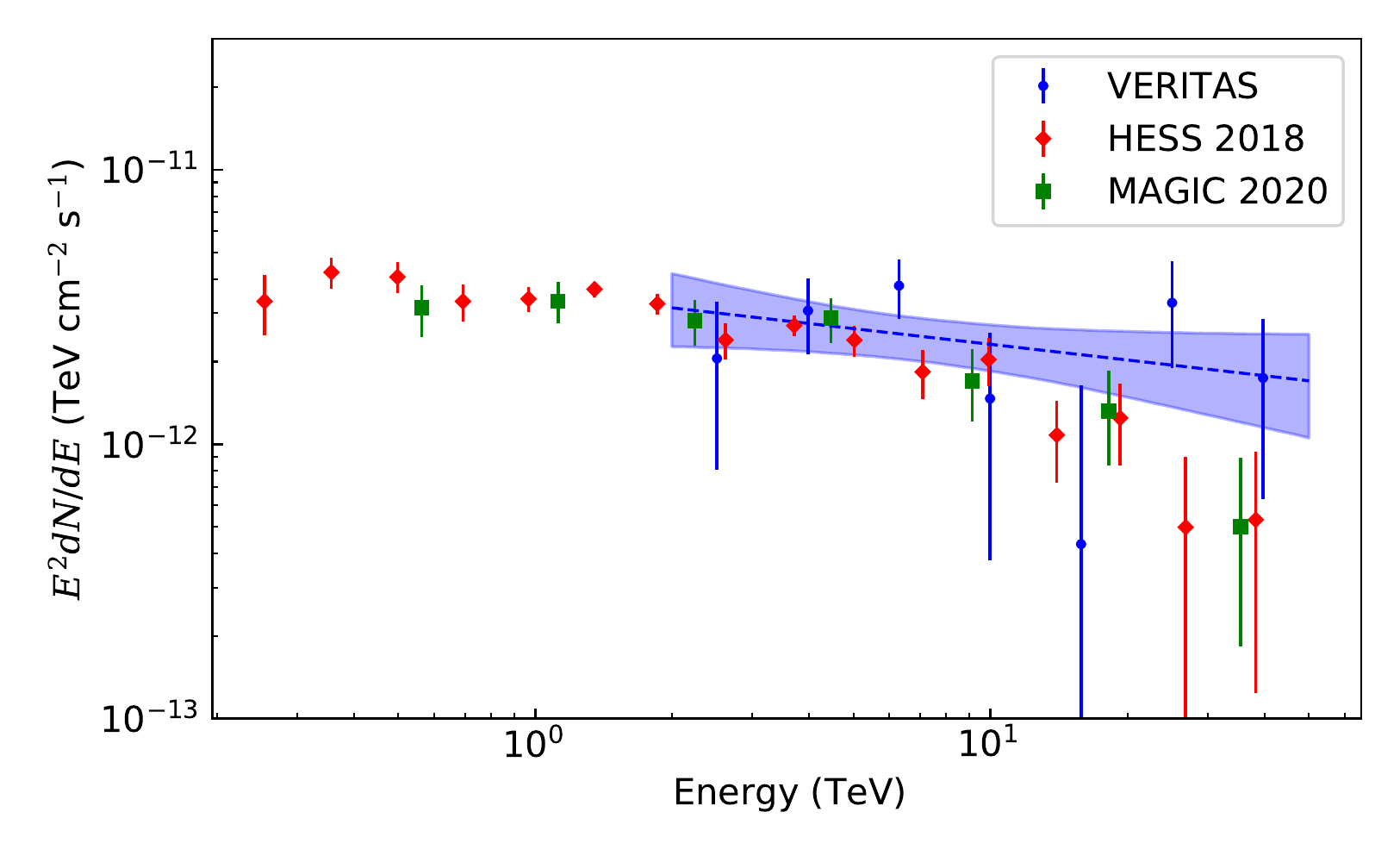}
  \caption
  {\edits{Differential energy spectrum of the diffuse ridge emission measured by VERITAS (blue) from the combined circular regions along the GC ridge (blue), with the normalization scaled to what would be measured in the signal region used by \citep{Abdalla2018}, as described in Section \ref{sec:results_diffuse_spectrum}. For comparison, the spectra measured by H.E.S.S. \citep[red;][]{Abdalla2018} and MAGIC \citep[green;][]{Acciari2020} are shown.
  Error bars represent $1\sigma$ uncertainties.
  The best-fit power law (blue dashed line) and $1\sigma$ confidence band on the model fit (shaded region) are also shown.}
  \label{fig:diffuse_spectrum_ver_hess}}
\end{figure*}

\edits{The H.E.S.S. \citep{Abdalla2018} and MAGIC \citep{Acciari2020} measurements of the diffuse ridge spectrum model larger regions than what we use in our analysis.
In order to compare our spectra, we can scale up our spectrum to estimate what would be measured in the signal region used by \citet{Abdalla2018}; the flux in this region is also comparable to the measurement of \citet{Acciari2020}.
Using the same 2D map of the diffuse emission used in the contamination calculation, we calculate the expected flux in the rectangular \citet{Abdalla2018} signal region, and in the regions used in the present analysis, finding a ratio of 2.7.
The scaled diffuse ridge emission differential energy spectrum (that is, the spectrum we measure times 2.7) is shown in \autoref{fig:diffuse_spectrum_ver_hess}, with the H.E.S.S. \citep{Abdalla2018} and MAGIC \citep{Acciari2020} spectra.}

The spectral index found in this analysis is consistent with the H.E.S.S. index of \statsys{2.28}{0.03}{0.2} \citep{Abdalla2018} \edits{and MAGIC index of ${1.98^{+0.21}_{-0.26\;\text{stat}}}{^{+0.15}_{-0.16\;\text{sys}}}$}, which supports the assumption that the diffuse emission across the ridge all comes from a common mechanism.
The index is also harder than the cosmic ray spectrum at Earth, which has an index of about 2.6 to 2.7 at these same energies \citep{Tanabashi2018}, suggesting that the cosmic ray spectrum near the GC is harder than that measured near the Earth.
The lack of a significant cut-off in the diffuse ridge emission spectrum is consistent with the presence of a PeVatron \citet{Abramowski2016},
\edits{though a cutoff below 1 PeV is not excluded by our data.
Using the parameterization and assumptions of \citet{Kelner2008}, we find that the proton spectrum providing the best fit to our gamma-ray spectrum is a PL with index $\sim 2.3$, and that for an ECPL proton spectrum with spectral index equal to 2.3, the 95\% lower limit on the cutoff energy is 0.08 PeV.
}
This index is consistent with \citet{Abramowski2016}, who found that their spectrum was best fit using a power-law proton spectrum with index $\sim 2.4$.

With greater statistics, the relationship between the spectral index and distance from the Galactic Center could be studied, which could give an indication of the composition of cosmic rays responsible for the diffuse emission.
For example, if protons are primarily responsible for the diffuse emission, a lack of softening in the spectrum as distance from the Galactic Center increases would be observed, as high-energy protons do not lose their energy as fast as electrons during diffusion.
Investigations into any potential variability of the flux of diffuse emission could also provide more information about the emission mechanism and possible past activity in the region.

\section{Other Sources in the CMZ}
In this VERITAS analysis, we studied multiple additional point sources in the CMZ.
These include the SNR G0.9+0.1, HESS J1746--285, and H.E.S.S. sources J1745--303 and J1741--302.

\subsection{SNR G0.9+0.1} \label{sec_results_g09}
\begin{figure}[!t]
 \centering
  \includegraphics[width=0.45\textwidth]{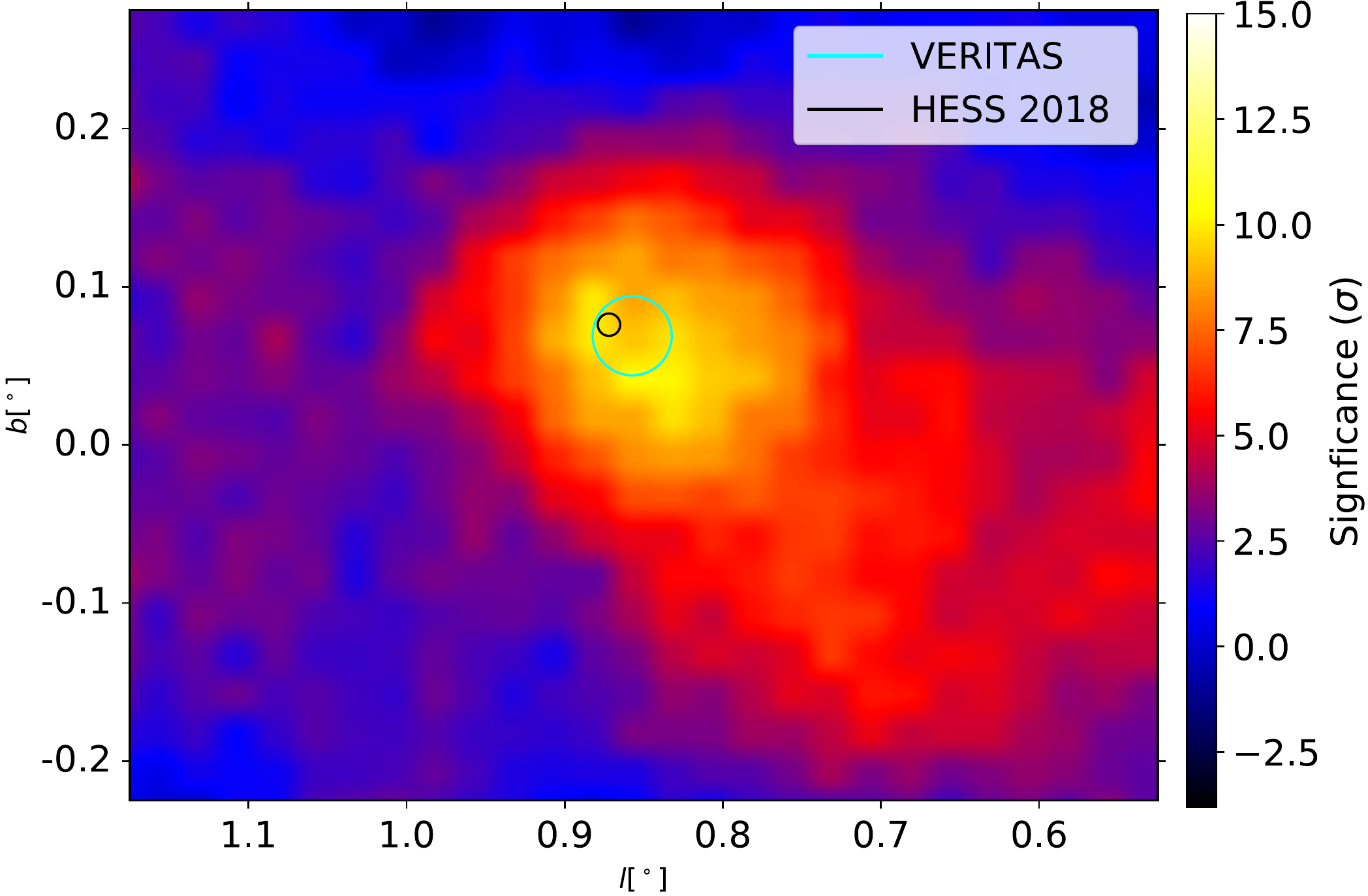}
  \caption{Significance map of the region around G0.9+0.1, showing gamma-ray significances as determined by VERITAS in this analysis.
  The best-fit positions and 68\% confidence regions are shown for this work (cyan) and H.E.S.S. \citep[black;][]{HGPS}.
  } \label{fig:g09_position_uncertainty}
\end{figure}

The second brightest VHE point source we identify in the GC region is VER J1747--281, which is associated with the composite SNR G0.9+0.1.
A close-up significance map of G0.9+0.1 is shown in \autoref{fig:g09_position_uncertainty}.
G0.9+0.1 is detected with a significance of $8\sigma$ at the best-fit position of \galcoords{0.857}{0.069}.
The statistical and systematic uncertainty on each coordinate are $0.03\degr$ and $0.013\degr$, respectively.

\begin{figure}[!t]
 \centering
  \includegraphics[width=0.475\textwidth]{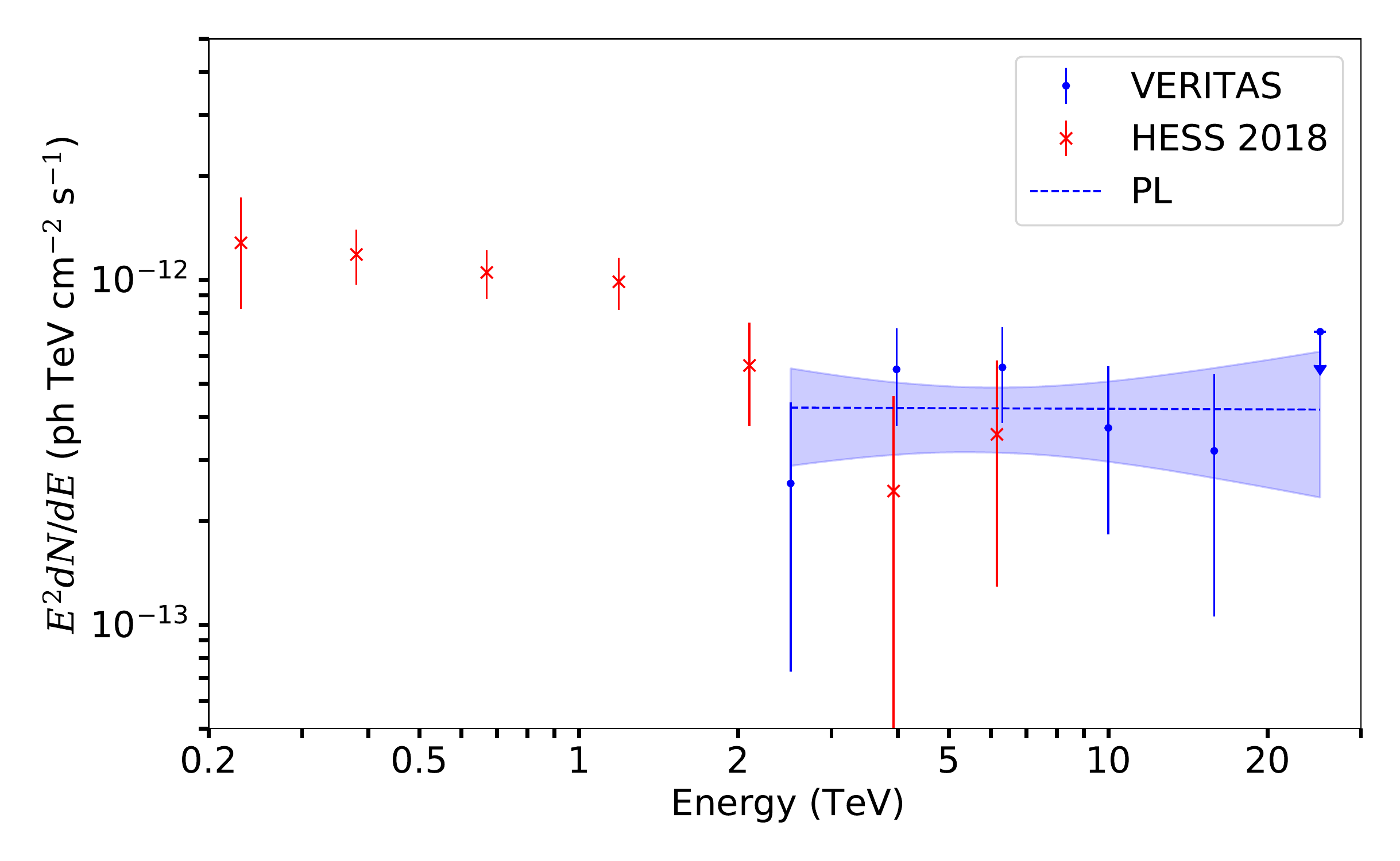}
  \caption{Differential energy spectrum of G0.9+0.1, as measured by VERITAS (blue), with the H.E.S.S. spectrum also shown \citep[red;][]{HGPS}.
  Error bars represent $1\sigma$ uncertainties in the flux.
  The downward arrow represents a 95\% upper limit.
  The best-fit power law (blue dashed line) is shown.
  The shaded region represents the $1\sigma$ confidence band on the model fit.}
  \label{fig:g09_spectrum}
\end{figure}

The differential energy spectrum of SNR G0.9+0.1 is shown in \autoref{fig:g09_spectrum}, with the H.E.S.S spectrum \citep{HGPS}.
The best fit to the joint spectrum is a PL with a spectral index of $2.00 \pm 0.28$ and flux normalization of \fluxnorm{1.51}{0.30}{-14} at 5.3 TeV.
The PL provides a good fit, with a $\chi^2$/d.o.f. of 2.28/3, corresponding to a $p$-value of 0.52.

\subsection{HESS J1746--285}
\citet{Archer2016} identified an excess in their significance map adjacent to VER J1745--290 which they named VER J1746--289, indicated in \autoref{fig:gc_sigmap_full_fov}.
The position of VER J1746--289 is inconsistent with the positions of nearby sources HESS J1746--285 \citep{Abdalla2018} and MAGIC J1746.4--2853 \citep{Ahnen2017}.
\edits{The position reported by \citet{Archer2016} is biased due to not accounting for the diffuse ridge emission, so we proceed using the position of HESS J1746--285, due to the more detailed modeling done by \citet{Abdalla2018}.}

\begin{figure}[!t] \centering
  \includegraphics[width=0.45\textwidth]
  {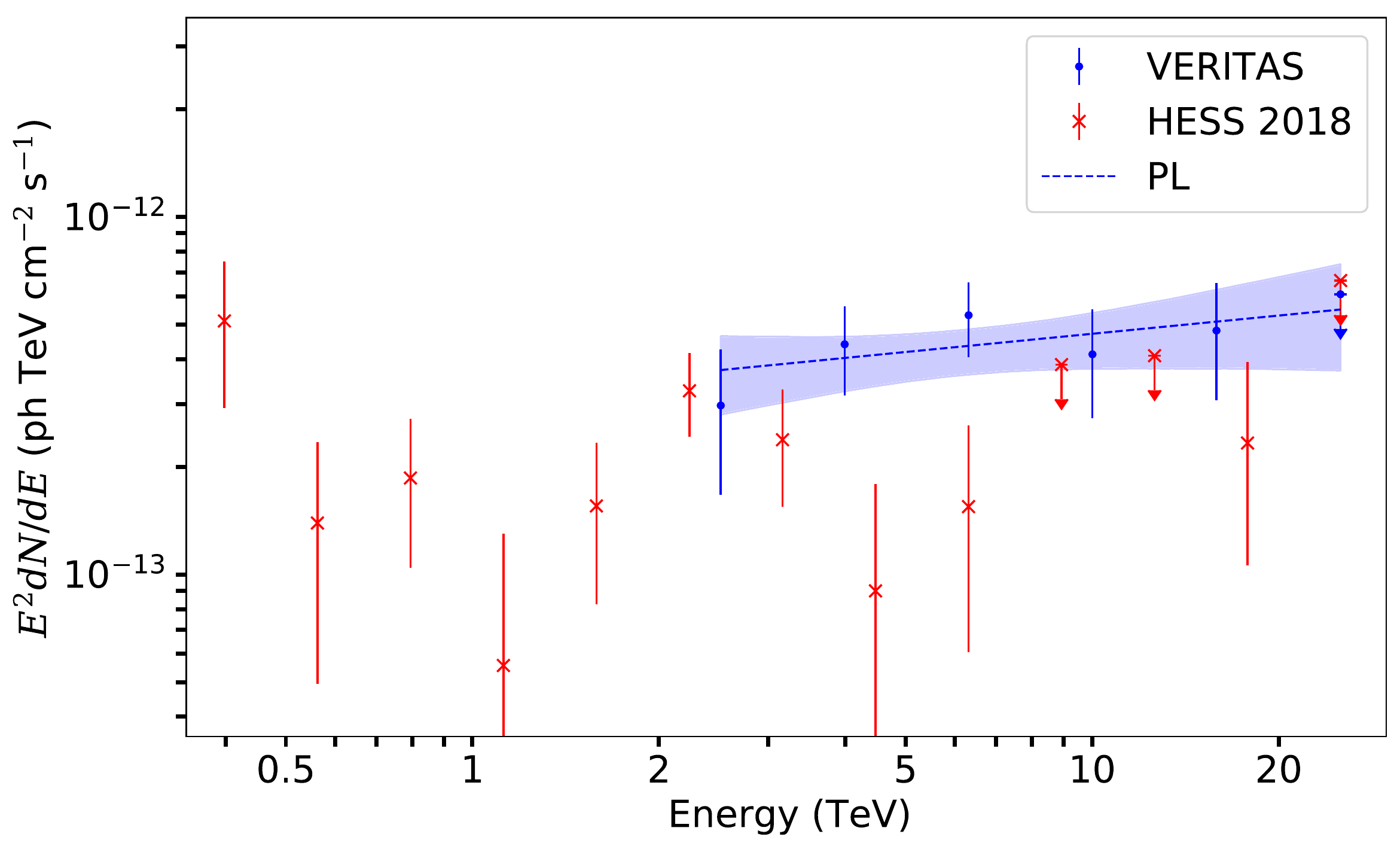}
  \caption{Differential energy spectrum at the position of HESS J1746--285 \citet{Abdalla2018}, as measured by VERITAS (blue), with the H.E.S.S. spectrum also shown \citep[red;][]{HGPS}.
  Error bars represent $1\sigma$ uncertainties in the flux.
  The downward arrow represents a 95\% upper limit.
  The best-fit power law (blue dashed line) is shown.
  The shaded region represents the $1\sigma$ confidence band on the model fit.}
  \label{fig:j1746-289_spectrum}
\end{figure}

The differential energy spectrum extracted from a $0.1\degr$ region around the position of HESS J1746--285 is shown in \autoref{fig:j1746-289_spectrum}, with the corresponding intrinsic spectrum from \citet{Abdalla2018}.
Our spectrum includes contributions from the central source and the diffuse emission.
The best fit to the joint spectrum is a PL with a spectral index of $1.83 \pm 0.22$ and flux normalization of \fluxnorm{1.51}{0.22}{-14} at 5.3 TeV.
The PL provides an adequate fit, with a $\chi^2$/d.o.f. of 1.21/3, corresponding to a $p$-value of 0.75.
H.E.S.S. and MAGIC find consistent, though softer spectral indices of $2.2\pm0.2$ and $2.29^{+0.19}_{-0.17,\text{stat}}\,^{+0.13}_{-0.23,\text{\,sys}}$, respectively \citep{Abdalla2018,Acciari2020}.
\edits{Following a similar procedure to the diffuse ridge emission contamination calculation, we find that other sources have an estimated contribution to our total integrated flux measurement of HESS J1746--285 of approximately 50\%, inhibiting comparisons to the H.E.S.S. and MAGIC results.}

The source's position coincides with a \fermilat{} source \citep[4FGL J1746.4-2852;][]{Abdollahi2020} and the Galactic radio arc \citep{Yusef-Zadeh1984,Yusef-Zadeh2004}, as noted by \citet{Archer2016}.
An extrapolation of the log-parabolic spectrum of 4FGL J1746.4-2852 lies well below our VHE spectrum, suggesting a different origin of the emission, even if the sources are associated.
\citet{Pohl1997} proposed that inverse-Compton scattering of far infrared radiation by electrons in the radio arc may be responsible for the MeV to GeV emission, and this scenario may produce TeV emission as well.
\citep{Abdalla2018} consider PWN candidate G0.13--0.11 \citep{Wang2002} to be the most likely counterpart.

\subsection{HESS J1741--302 and HESS J1745--303}
\label{sec:300}

The extended H.E.S.S. sources HESS J1745--303 and HESS J1741--302 fall within the FoV of VERITAS GC observations.
HESS J1745--303 is located at Galactic coordinates \galcoords{358.71}{-0.64}, which is near in proximity to the shell-like SNR G359.1--0.5.
Emission from J1745$-$303 is consistent with the model in which the particles accelerated to high energies by the SNR interact with dense matter.
While some excess signal is seen by VERITAS near HESS J1745--303 (with a statistical significance of about $4\sigma$), it is not strong enough to usefully characterize the source.
Little excess is seen in the HESS J1741--302 region, however the acceptance this far from Sgr A* is around half of the maximum acceptance.

\section{Summary and Conclusions}
\begin{deluxetable*}{| c | p{3cm} | p{4cm} | p{5.6cm} |}[ht]
  \tablehead{ \colhead{Source} & \colhead{Best Spectral Fit} & \colhead{Fit Parameters} & \colhead{Conclusions} }
\tablewidth{0.2\textwidth}
\tablecaption{Summary of the VERITAS results for the strongest VHE gamma-ray sources in the CMZ.
\label{tab:summary}}
\startdata
VER J1745--290	&	power law with	&	$\Gamma=2.12^{+0.22}_{-0.17}$	&	Consistent with PWN because of \\
	&	exponential cutoff	&	$E_\mathrm{cut}=10.0^{+4.0}_{-2.0}$ TeV	&	cutoff and lack of variability \\ \hline
Diffuse emission	&	power law	&	$\Gamma=2.19\pm0.20$	&	Consistent with a hadronic \newline accelerator to PeV energies \\ \hline
SNR G0.9+0.1	&	power law	&	$\Gamma=2.00 \pm 0.28$	&	Consistent with SNR \\ \hline
HESS J1746--285	&	power law	&	$\Gamma = 1.83 \pm 0.22$	&	Unidentified \\
\enddata
\end{deluxetable*}

A summary of results from this analysis is shown in \autoref{tab:summary}.
The results are a major improvement over previous VERITAS papers on the GC region due to the use of additional data and advanced analysis techniques.  
The estimated positions of the VHE point sources detected by VERITAS in this analysis are consistent with previous results, from both VERITAS and H.E.S.S. 
The positional uncertainty of the central source VER J1745--290 is too large to exclude Sgr A East or G359.95--0.04 as counterparts to the source.

Increased statistics on the high end of the energy spectrum  of VER J1745--290 show evidence of a cutoff or break energy around 10 TeV, consistent with H.E.S.S. and MAGIC \citet{Abdalla2018,Acciari2020}.
A cutoff is predicted by leptonic models such as inverse-Compton models \citep[e.g.][]{Aharonian2005a}, plerion models \citep[e.g.][]{Atoyan2004,Kusunose2012}, \edits{and pair absorption, which is especially significant $\gtrsim$10 TeV \citep{Porter2018}}.
A cutoff at this energy is also consistent with the emission of SNRs and PWNe, most of which cut off between 10 and 20 TeV.
The low value of the cutoff energy could possibly indicate that the magnetic fields in this source are very strong causing a high rate of electron cooling. 

No variability of the flux of VER J1745--290 was observed for timescales of days to years.
\edits{PWNe are not known to exhibit variable behavior, so are consistent with the data, although the weak constraints on variability due to the low counting statistics do not allow for strong conclusions concerning the nature of the source}.

We have produced the first spectrum of the diffuse emission in the GC region as determined by VERITAS.
The diffuse emission observed by VERITAS, in contrast to the central source VER J1745--290, does not show any evidence of a cutoff, maintaining a hard power law up to 40 TeV.
The lack of a cutoff in the spectrum supports the case for the presence of an accelerator of PeV cosmic ray particles in the GC.

The total energy for the flux of radiation in the region could have been supplied by a large supernova event in the past.
The energy flux is also consistent with a population of $10^4$ to $10^5$ millisecond pulsars with luminosities above $10^{34}$\,erg\,s$^{-1}$ and modest acceleration efficiencies scattered throughout the CMZ \citep{Guepin2018}.
Further observations by a VHE gamma-ray telescope with an angular resolution better than VERITAS should be able to provide evidence for, or against, steady-state hadronic models based on the observed morphology of the TeV emission.

The major new worldwide project in the VHE gamma-ray waveband is the Cherenkov Telescope Array \citep[CTA;][]{Acharya2018}.
CTA will offer substantially improved sensitivity, a wider energy range and improved angular resolution (below $0.05\degr$ above 1 TeV) compared to the existing IACTs.
In addition, CTA is expected to have a systematic pointing uncertainty of less than $10\arcsec$ which could be sufficiently precise to distinguish between Sgr A* and PWN G395.95--0.04 as the source of gamma rays from J1745--290.

Pushing the spectral reconstruction of both J1745--290 and the diffuse component to higher energies will be important to better determine the relevant acceleration processes. 
With its much larger collection area, CTA will have an energy reach out to 100\,TeV and beyond. 
Flares, if they occur in the VHE band, will be much more likely to be detected by CTA with its greatly improved sensitivity.

A spatial correlation between TeV neutrinos and diffuse gamma rays could strengthen the hadronic scenario of diffuse emission in the CMZ.
Data from experiments such as \textit{IceCube} \citep{Aartsen2017} could possibly produce such a result in the future.

\section{Acknowledgments}
This research is supported by grants from the U.S. Department of Energy Office of Science, the U.S. National Science Foundation and the Smithsonian Institution, by NSERC in Canada, and by the Helmholtz Association in Germany. This research used resources provided by the Open Science Grid, which is supported by the National Science Foundation and the U.S. Department of Energy's Office of Science, and resources of the National Energy Research Scientific Computing Center (NERSC), a U.S. Department of Energy Office of Science User Facility operated under Contract No. DE-AC02-05CH11231. We acknowledge the excellent work of the technical support staff at the Fred Lawrence Whipple Observatory and at the collaborating institutions in the construction and operation of the instrument.

A portion of our support came from these awards from the National Science Foundation:
PHY-1307171, ``Particle Astrophysics with VERITAS and Defining Scientific Horizons for CTA"
and
PHY-1607491, ``Particle Astrophysics with VERITAS and Development for CTA."

\appendix

\setcounter{table}{0}
\renewcommand{\thetable}{A\arabic{table}}

\begin{deluxetable*}{cccc}[ht]
   \tablehead{
\colhead{Energy}	&	\colhead{Flux}	&	\colhead{Flux Uncertainty}	&	\colhead{Significance}	\\
\colhead{(TeV)}	&	\colhead{(\fluxunits)}	&	\colhead{(\fluxunits)}	&	\colhead{($\sigma$)}
} 
\tablecaption
{Differential flux of VER J1745--290 measured between 2010 April and 2018 June by VERITAS. Flux uncertainties are 1$\sigma$ statistical uncertainties. Fluxes with no associated uncertainties are 95\% upper limits.}
\label{tab:spec_points_j1745}
\startdata
2.11	&	$7.7\times 10^{-13}$	&	$0.9\times 10^{-13}$	&	13.29	\\
2.66	&	$4.0\times 10^{-13}$	&	$0.5\times 10^{-13}$	&	15.15	\\
3.35	&	$2.3\times 10^{-13}$	&	$0.3\times 10^{-13}$	&	13.61	\\
4.21	&	$1.6\times 10^{-13}$	&	$0.2\times 10^{-13}$	&	15.34	\\
5.30	&	$6.8\times 10^{-14}$	&	$0.9\times 10^{-14}$	&	11.37	\\
6.68	&	$3.6\times 10^{-14}$	&	$0.5\times 10^{-14}$	&	10.30	\\
8.40	&	$1.9\times 10^{-14}$	&	$0.3\times 10^{-14}$	&	 8.01	\\
10.6	&	$1.2\times 10^{-14}$	&	$0.2\times 10^{-14}$	&	 8.22	\\
13.3	&	$6.3\times 10^{-15}$	&	$1.3\times 10^{-15}$	&	 6.53	\\
16.7	&	$2.1\times 10^{-15}$	&	$0.7\times 10^{-15}$	&	 3.91	\\
21.6	&	$7.5\times 10^{-16}$	&	$3.1\times 10^{-16}$	&	 3.24	\\
30.9	&	$9.2\times 10^{-17}$	&	$7.2\times 10^{-17}$	&	 1.65	\\
48.4	&	$9.4\times 10^{-17}$	&	\text{}	&	-0.48	\\
\enddata
\end{deluxetable*}

\begin{deluxetable*}{cccc}[ht]
\tablehead{
\colhead{Energy}	&	\colhead{Flux}	&	\colhead{Flux Uncertainty}	&	\colhead{Significance}	\\
\colhead{(TeV)}	&	\colhead{(\fluxunits)}	&	\colhead{(\fluxunits)}	&	\colhead{($\sigma$)}
}
\tablecaption
{{Differential flux of the diffuse Galactic ridge measured between 2010 April and 2018 June by VERITAS. Flux uncertainties are 1$\sigma$ statistical uncertainties.}}
\label{tab:spec_points_diffuse}
\startdata
2.51	&	$1.2\times 10^{-13}$	&	$7.4\times 10^{-14}$	&	7.60	\\
3.97	&	$7.2\times 10^{-14}$	&	$2.2\times 10^{-14}$	&	7.46	\\
6.30	&	$3.5\times 10^{-14}$	&	$0.9\times 10^{-14}$	&	7.44	\\
9.98	&	$5.5\times 10^{-15}$	&	$4.1\times 10^{-15}$	&	3.31	\\
15.8	&	$6.4\times 10^{-16}$	&	$17.8\times 10^{-16}$	&	1.77	\\
25.1	&	$1.9\times 10^{-15}$	&	$8.1\times 10^{-16}$	&	3.71	\\
39.7	&	$4.1\times 10^{-16}$	&	$2.6\times 10^{-16}$	&	2.24	\\
\enddata
\end{deluxetable*}

\bibliography{references}
\bibliographystyle{apj}

\end{document}